\documentclass[seceq]{ptptex}

\usepackage{graphicx}

\title{
Yukawa's Pion, Low-Energy QCD and Nuclear Chiral Dynamics%
}

\author{
Wolfram \textsc{Weise}%
}

\inst{
Physik-Department, Technische Universit\"at M\"unchen, D-85747 Garching, Germany
}

\abst{
A survey is given of the evolution from Yukawa's early work, via the understanding of the pion as a Nambu-Goldstone boson of spontaneously broken chiral symmetry in QCD, to modern developments in the theory of the nucleus based on the chiral effective field theory representing QCD in its low-energy limit.}

\begin{document}

\maketitle

\section{The Beginnings}

One of the most remarkable documentations in modern science history is the series of articles which form the first volumes of Supplement of the Progress of Theoretical Physics. These Collected Papers on Meson Theory \cite{Sup55} celebrated the twenty years' anniversary of Yukawa's pioneering work \cite{Yuk35} which laid the foundations for our understanding of the strong force between nucleons. In those early days of nuclear physics very little was known about this force. Heisenberg's "Platzwechsel" interaction between proton and neutron gave a first intuitive picture, and 
some unsuccessful attempts had been made to connect Fermi's theory of beta decay with nuclear interactions. 

Yukawa's 1935 article introduced the conceptual framework for a systematic approach to the nucleon-nucleon interaction. His then postulated "$U$ field" did not yet have the correct quantum numbers of what later became the pion\footnote{I am grateful to Professor H. Miyazawa for instructive communications on this point.}, but it had already some of its principal features. It provided the basic mechanism of charge exchange between proton and neutron. From the form of the potential, $e^{-\mu r}/r$, and from the estimated range of the nuclear force,  the mass of the $U$ particle was predicted to be about 200 times that of the electron. This $U$ particle was erroneously identified with the muon discovered in 1937 almost simultaneously by Anderson and by Nishina and their collaborators. The "real" $\pi$ meson was discovered a decade later\cite{pion47} in cosmic rays and then produced for the first time\cite{pion48} at the Berkeley cyclotron. Its pseudoscalar nature was soon established. In 1949 the Nobel prize was awarded to Yukawa.

\begin{figure}
       \centerline{\includegraphics[width=5.5cm] {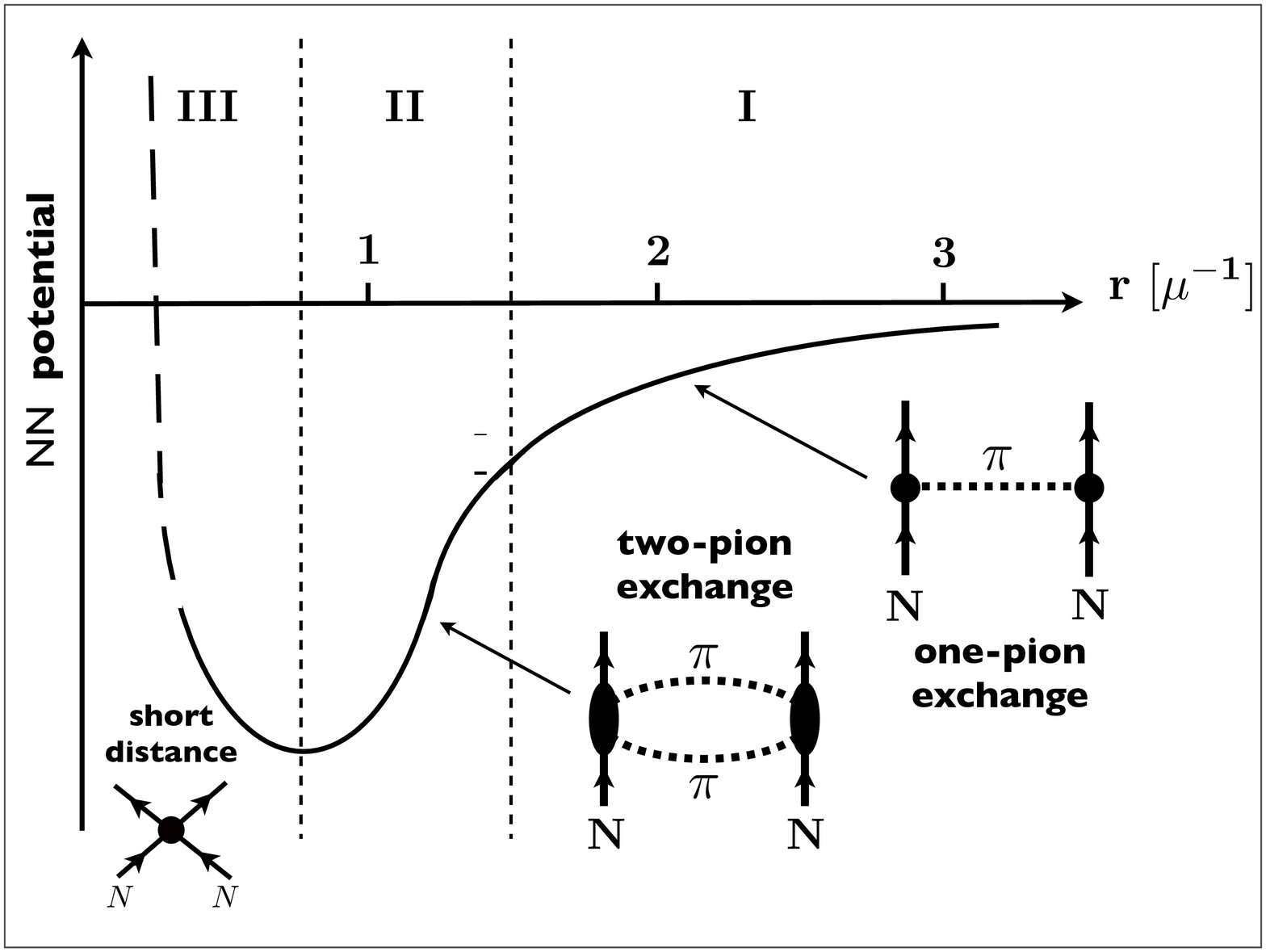}}
   \caption{Hierarchy of scales governing the nucleon-nucleon potential (adapted from Taketani  \cite{Tak56}). \newline The distance $r$ is given in units of the pion Compton wavelength $\mu^{-1} \simeq 1.4$ fm.}
   \label{fig:1}
 \end{figure}

In the footsteps of Yukawa's original work, the decade thereafter saw an impressively coherent effort by the next generation of Japanese theorists.  A cornerstone of these developments was the visionary conceptual design by Taketani et al. \cite{TNS51} of the inward-bound hierarchy of scales governing the nucleon-nucleon interaction, sketched in Fig.\ref{fig:1}. The long distance region I 
is determined by one-pion exchange. It continues inward to the intermediate distance region II dominated by two-pion exchange. The basic idea was to construct the NN potential in regions I and II by explicit calculation of $\pi$ and $2\pi$ exchange processes, whereas the detailed behaviour of the interaction in the short distance region III remains unresolved at the low-energy scales characteristic of nuclear physics. This short distance part is given a suitably parametrized form. The parameters are fixed by comparison with scattering data. 
    
Taketani's picture, although not at all universally accepted at the time by the international community of theorists, turned out to be immensely useful. Today this strategy is the one conducted by modern effective field theory approaches. It is amazing how far this program had already been developed by Japanese theorists in the late fifties. One example is the pioneering calculation of the two-pion exchange potential \cite{KMO57} using dispersion relation techniques and early knowledge \cite{FM50} of the resonant pion-nucleon amplitude which anticipated the $\Delta$ isobar models of later decades.

\section{The Pion in the context of Low-Energy QCD}

Today's theory of the strong interaction is Quantum Chromodynamics (QCD). 
There exist two limiting situations in which QCD is accessible with "controlled" approximations. At momentum scales exceeding several GeV (corresponding to short distances, $r < 0.1$ fm), QCD is a theory of weakly interacting quarks and gluons. At low momentum scales considerably smaller than 1 GeV (corresponding to long distances, $r > 1$ fm), QCD is governed by color confinement and a non-trivial vacuum: the ground state of QCD hosts strong condensates of quark-antiquark pairs and gluons.
Confinement implies the dynamical breaking of a global symmetry which is exact in the limit of massless quarks: chiral symmetry. Spontaneous chiral symmetry breaking in turn implies the existence of pseudoscalar Nambu-Goldstone bosons. For two quark flavours $(N_f = 2)$ with (almost) massless $u$ and $d$ quarks, these Goldstone bosons are identified with the isospin triplet of pions. At low energy, Goldstone bosons interact weakly with one another or with any massive particles. Low-energy QCD is thus realized as an effective field theory in which these Goldstone bosons are the active, light degrees of freedom.

\subsection{Chiral Symmetry and the NJL Model}
How does Yukawa's pion figure in the frame of QCD? Historically, the foundations for understanding the pion as a Nambu-Goldstone boson \cite{G61,NJL61} of spontaneously broken chiral symmetry were initiated in the 1960's, culminating in the PCAC and current algebra approaches\cite{AD68} of the pre-QCD era. A most inspiring work from this early period is the one by Nambu and Jona-Lasinio\cite{NJL61}(NJL). Just as the BCS model provided an understanding of the basic mechanism behind superconductivity, the NJL model helped clarifying the dynamics that drives spontaneous chiral symmetry breaking and the formation of pions as Goldstone bosons. 

Consider as a starting point the conserved color current of quarks, ${\bf J}_\mu^a = \bar{\psi}\gamma_\mu {\bf t}^a\psi$, where ${\bf t}^a$ ($a = 1, ... ,8$) are the generators of the $SU(N_c=3)$ gauge group and $\psi$ are the quark fields with $4N_c N_f$ components representing their spin, color and flavor degrees freedom. This current couples to the gluon fields. Any two quarks interact through multiple exchanges of gluons. Assume that the distance over which color propagates is restricted to a short correlation length $l_c$. Then the interaction experienced by low-momentum quarks can be schematically viewed as a local coupling between their color currents:
\begin{equation}
{\cal L}_{int} = -G_c\,{\bf J}_\mu^a(x)\,{\bf J}^\mu_a(x)\,\, ,
\label{eq:Lint1}
\end{equation}
where $G_c \sim \bar{g}^2\, l_c^2$ is an effective coupling strength of dimension $length^2$ which encodes the QCD coupling, averaged over the relevant distance scales, in combination with the squared correlation length, $l_c^2$. 

Now adopt the local interaction (\ref{eq:Lint1}) and write the following model Lagrangian for the quark fields $\psi(x)$:
\begin{equation}
{\cal L} = \bar{\psi}(x)(i\gamma^\mu\partial_\mu - m_0)\psi(x) + {\cal L}_{int}(\bar{\psi},\psi)\,\, .
\label{eq:NJL}
\end{equation}
In essence, by "integrating out" gluon degrees of freedom and absorbing them in the four-fermion interaction ${\cal L}_{int}$, the local $SU(N_c)$ gauge symmetry of QCD is now replaced by a global $SU(N_c)$ symmetry. Confinement is obviously lost, but all other symmetries of QCD are maintained. The mass matrix $m_0$ incorporates small "bare" quark masses. In the limit $m_0 \rightarrow 0$, the Lagrangian (\ref{eq:NJL}) has a chiral symmetry of right- and left-handed quarks, $SU(N_f)_R\times SU(N_f)_L$, that it shares with the original QCD Lagrangian for $N_f$ massless quark flavors.

A Fierz transform of the color current-current interaction (\ref{eq:Lint1}) produces a set of exchange terms
acting in quark-antiquark channels. For the $N_f = 2$ case: 
\begin{equation}
{\cal L}_{int} \rightarrow {G\over 2}\left[(\bar{\psi}\psi)^2 + (\bar{\psi}\,i\gamma_5\,\mbox{\boldmath$\tau$}\,\psi)^2\right] + ... \,\, , 
\label{eq:Lint2}
\end{equation}
with the isospin $SU(2)$ Pauli matrices $\mbox{\boldmath$\tau$} = (\tau_1,\tau_2,\tau_3)$. For brevity we have not shown a series of terms with combinations of vector and axial vector currents, both in color singlet and color octet channels. The constant $G$ is proportional to the color coupling strength $G_c$. The ratio of these two constants is uniquely determined by $N_c$ and $N_f$.

The steps just outlined can be viewed as a contemporary way of introducing the time-honored NJL model \cite{NJL61} . This model has been further developed and applied \cite{VW91,HK94} to a variety of problems in hadron physics. The virtue of this schematic model is its simplicity in illustrating the basic mechanism of spontaneous chiral symmetry breaking, as follows. In the mean-field (Hartree) approximation the equation of motion derived from the Lagrangian (\ref{eq:NJL}) leads to a gap equation
\begin{equation}
M = m_0 - G\langle\bar{\psi}\psi\rangle\,\, ,
\end{equation}
which links the dynamical generation of a constituent quark mass $M$ to the appearance of the chiral quark condensate
\begin{equation}
\langle\bar{\psi}\psi\rangle= -Tr\lim_{\,x\rightarrow \,0^+}\langle {\cal T}\psi(0)\bar{\psi}(x)\rangle = -2iN_fN_c\int {d^4p\over (2\pi)^4}{M\,\theta(\Lambda^2 -\vec{p}\,^2)\over p^2 - M^2 + i\varepsilon}\,\, .
\end{equation}
This condensate plays the role of an order parameter of spontaneous chiral symmetry breaking. Starting from $m_0 = 0$ a non-zero quark mass develops dynamically, together with a non-vanishing chiral condensate, once $G$ exceeds a critical value of order $G_{crit} \sim 10$ GeV$^{-2}$. The procedure requires a momentum cutoff $\Lambda \simeq 2M$ beyond which the interaction is "turned off". Note that the strong non-perturbative interactions, by polarizing the vacuum and turning it into a condensate of quark-antiquark pairs, transmute an initially pointlike quark with its small bare mass $m_0$ into a massive quasiparticle with a size of order $(2M)^{-1}$. 

\subsection{The Pseudoscalar Meson Spectrum}
The NJL model demonstrates lucidly the appearance of chiral Nambu-Goldstone bosons. Solving Bethe-Salpeter equations in the color singlet quark-antiquark channels generates the lightest mesons as quark-antiquark excitations of the correlated QCD ground state with its condensate structure. 
Several such calculations have been performed in the past with $N_f = 3$ quark flavors \cite{KLVW90,VW91,HK94} . Such a model has an unwanted $U(3)_R\times U(3)_L$ symmetry to start with, but the axial $U(1)_A$ anomaly reduces this symmetry to $SU(3)_R\times SU(3)_L\times U(1)_V$. In QCD, instantons are considered responsible for $U(1)_A$ breaking. In the NJL model, these instanton driven interactions are incorporated in the form of a flavor determinant \cite{tH76} $\det[\bar{\psi}_i(1 \pm \gamma_5)\psi_j]$. This interaction involves all three flavors $u, d, s$ simultaneously in a genuine three-body term. 

 \begin{figure}
       \centerline{\includegraphics[width=6.3cm] {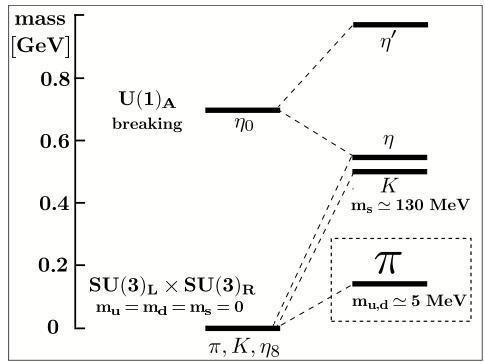}}
   \caption{Symmetry breaking pattern in the pseudoscalar meson nonet calculated in the three-flavor NJL model \cite{KLVW90} .}
   \label{fig:2}
 \end{figure}

The symmetry breaking pattern resulting from such a calculation is apparent in the pseudoscalar meson spectrum of Fig.\ref{fig:2}. Starting from massless $u, d$ and $s$ quarks, the pseudoscalar octet emerges as a set of massless Goldstone bosons of spontaneously broken $SU(3)\times SU(3)$, while $U(1)_A$ breaking drives the singlet $\eta_0$ away from the Goldstone boson sector. Finite quark masses shift the $J^\pi = 0^-$ nonet into its empirically observed position, including $\eta$-$\eta'$ mixing. 

The very special nature of the pion as a Nambu-Goldstone boson is manifest in a famous relation\cite{GOR68} derived from current algebra and PCAC:
\begin{equation}
m^2_{\pi}\,f_{\pi}^2 = - {1\over 2}(m_u + m_d) \langle \bar{\psi} \psi \rangle + 
{\cal O}(m^2_{u,d}) .
\label{eq:GOR}
\end{equation}
It involves the pion decay constant, $f_\pi \simeq 0.09$ GeV, defined by the matrix element which connects the pion with the QCD vacuum via the axial vector current, $A_i^\mu = \bar{\psi}\gamma^\mu\gamma_5{\tau_i\over 2} \psi$:
\begin{equation}
\langle 0 | A^{\mu}_i (x=0) | \pi_i (p) \rangle = i p^{\mu} f _\pi\, .
\label{eq:fpi}
\end{equation}
Just like the chiral condensate, the pion decay constant is a measure of spontaneous chiral symmetry breaking expressed in terms of a characteristic scale $4\pi f_\pi \sim 1$ GeV. The non-zero pion mass, $m_\pi \sim 0.14$ GeV $\ll 4\pi f_\pi$,
reflects the explicit symmetry breaking by the small quark masses,
with $m^2_{\pi} \sim m_q$. One should note that the quark masses $m_{u,d}$ and the condensate $\langle\bar{\psi} \psi \rangle$ are both scale dependent quantities. Only their product is scale independent, i.e. invariant under the renormalization group. At a renormalization scale of about 1 GeV, a typical average quark mass ${1\over 2}(m_u + m_d) \simeq 7$ MeV implies $|\langle\bar{\psi} \psi \rangle| \simeq $ (0.3 GeV)$^3$.

\subsection{Scales and Symmetry Breaking Patterns}

The quark masses are the only parameters that set primary scales in QCD. Their classification into sectors of "light" and "heavy" quarks determines very different physics phenomena. While the heavy quarks (i.e. the $t$, $b$ and - within limits - the $c$ quarks) offer a natural "small parameter" in terms of their reciprocal masses, such that non-relativistic approximations (expansions of observables in powers of $m_{t,b,c}^{-1}$) tend to work increasingly well with increasing quark mass,
the sector of the light quarks (i.e. the $u$, $d$ quarks and - to some extent - the $s$ quark) is governed by quite different principles and rules. Evidently, the quark masses themselves are now "small parameters", to be compared with a characteristic "large" scale of dynamical origin. This large scale is  visible as a characteristic mass gap of about 1 GeV which separates the QCD vacuum from almost all of its excitations, with the exception of the pseudoscalar meson octet of pions, kaons and the eta meson. This mass gap is in turn comparable to $4\pi f_\pi$, the scale associated with spontaneous chiral symmetry breaking in QCD. 

 \begin{figure}
       \centerline{\includegraphics[width=5cm] {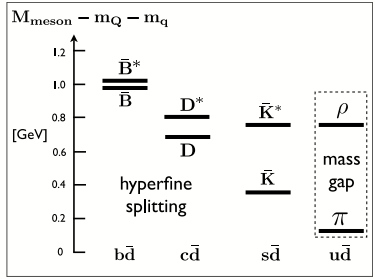}}
   \caption{ Evolution of the splitting between spin singlet (lower) and triplet (upper) quark-antiquark states (the pseudoscalar ($J^\pi = 0^-$) and vector ($J^\pi = 1^-$) mesons) with varying mass of one of the quarks. The bare quark masses are subtracted from the physical meson masses for convenience of demonstration. }
   \label{fig:3}
 \end{figure}

In this context it is also instructive to have a look at the spectroscopic pattern of pseudoscalar and vector mesons, starting from heavy-light quark-antiquark pairs in $^1S_0$ and $^3S_1$ states and following those states downward in the mass of the quark. This is illustrated in Fig.\ref{fig:3} where we show the masses of mesons composed of a $b$, $c$, $s$ or $u$ quark with an anti-$d$-quark attached. Bare quark masses are subtracted from the meson masses in this plot in order to directly demonstrate the evolution from perturbative hyperfine splitting in the heavy systems to the non-perturbative mass gap in the light ones. In the $\bar{B}$ and $\bar{B}^*$ mesons, the $\bar{d}$ quark is tightly bound to the heavy $b$ quark at small average distance, within the range where perturbative QCD is applicable. The spin-spin interaction is well approximated by perturbative one-gluon exchange, resulting in a small hyperfine splitting. Moving downward in mass to the $D$ and $D^*$ systems, with the $b$ quark replaced by a $c$ quark, the hyperfine splitting increases but remains perturbative in magnitude. As this pattern evolves further into the light-quark sector, it undergoes a qualitative change via the large mass difference of $\bar{K}$ and $\bar{K}^*$  to the non-perturbative mass gap in the $\pi - \rho$ system, reflecting the Goldstone boson nature of the pion.

\subsection{Chiral Effective Field Theory}

Low-energy QCD is the physics of systems of light quarks at energy and momentum scales smaller than the 1 GeV mass gap observed in the hadron spectrum. 
This 1 GeV scale set by $4 \pi f_{\pi} $ offers a natural separation between "light" and "heavy" (or, correspondingly, "fast" and "slow") degrees of freedom. The basic idea of an effective field theory is to
introduce the active light particles as collective degrees of freedom,  while the
heavy particles are frozen and treated as (almost) static sources. The dynamics
is described by an effective Lagrangian which incorporates all relevant
symmetries of the underlying fundamental theory. In QCD, confinement and spontaneous chiral symmetry breaking implies that the "fast" degrees of freedom are the Nambu-Goldstone bosons.
With Yukawa's pion in mind, we restrict ourselves to $N_f = 2$. 

We first briefly summarize the steps\cite{Wei67,GL84} required in the pure meson sector (baryon number $B$ = 0) and later for the pion-nucleon sector ($B$ = 1). A chiral field is introduced as 
\begin{equation}
U(x) = e^{i\, \tau_i\, \pi_i(x)/f_\pi} \in SU(2)~~,
\end{equation}
with the Goldstone pion fields $\pi_i(x)$ normalized by the pion decay constant $f_\pi$ taken in the chiral limit ($m_\pi \rightarrow 0$).
The QCD Lagrangian is replaced by an effective Lagrangian which
involves $U(x)$ and its derivatives:
\begin{equation}
{\cal L}_{QCD} \to {\cal L}_{eff} (U, \partial^\mu U, ...) .
\label{eq:Leff}
\end{equation}
Goldstone bosons interact only when they carry non-zero momentum, so the low-energy
expansion of (\ref{eq:Leff}) is an ordering in powers of $\partial^{\mu} U$. Lorentz
invariance permits only even numbers of derivatives. One writes
\begin{equation}
{\cal L}_{eff} = {\cal L}^{(2)} + {\cal L}^{(4)} + ... ~~,
\end{equation}
omitting an irrelevant constant. The leading term (the non-linear sigma
model) involves two derivatives:
\begin{equation}
{\cal L}^{(2)} = \frac{f_\pi^2}{4} Tr [ \partial_{\mu} U^{\dagger} \partial^{\mu} U
] .
\end{equation}
At fourth order, the terms permitted by symmetries are (apart from an extra
contribution from the QCD anomaly, not included here):
\begin{equation}
{\cal L}^{(4)} = \frac{l_1}{4} ( Tr [\partial_{\mu} U^{\dagger}
\partial^{\mu} U])^2 +
\frac{l_2}{4} Tr [\partial_{\mu} U^{\dagger} \partial_{\nu} U] Tr
[\partial^{\mu} U^{\dagger} \partial^{\nu} U] ,
\end{equation}
and so forth. The constants $l_1, l_2$ (following canonical notations\cite{GL84})
must be determined by experiment. 

The symmetry breaking mass term is small, so that it can be handled
perturbatively, together with the power series in momentum. The leading
contribution introduces a term linear in the quark mass matrix $m = diag(m_u,m_d)$:
\begin{equation}
{\cal L}^{(2)} = \frac{f_\pi^2}{4} Tr [\partial_{\mu} U^{\dagger} \partial^{\mu} U]
+ \frac{f_\pi^2}{2} B_0 \, Tr [m (U + U^{\dagger})] ,
\end{equation}
with $B_0 = -\langle\bar{\psi}\psi\rangle / 2f_\pi$
The fourth order term ${\cal L}^{(4)}$ also receives symmetry breaking contributions with additional constants $l_i$.

To the extent that the
effective Lagrangian includes all terms dictated by symmetries of QCD, the chiral
effective field theory is the low-energy equivalent\cite{Wei79,L94} of the original QCD Lagrangian.
Given the effective Lagrangian, the framework for systematic perturbative
calculations of the S-matrix involving Goldstone bosons, Chiral
Perturbation Theory (ChPT), is then defined by the following rules:\\ 
Collect all Feynman diagrams generated by ${\cal L}_{eff}$. Classify
all terms according to powers of a small quantity $Q$
which stands generically for three-momentum or energy 
of the Goldstone boson, or for the pion mass $m_{\pi}$. The small
expansion parameter is $Q/4\pi f_{\pi}$. Loops are subject
to dimensional regularization and renormalization. 
 
\subsection{Pion-Pion Scattering}

When using the Gell-Mann, Oakes,  Renner (GOR) relation (\ref{eq:GOR}) to leading order
in the quark mass, it is tacitly assumed that the chiral condensate is
large in magnitude and plays the role of an order parameter for
spontaneous chiral symmetry breaking. This basic scenario needs to be 
confirmed. It has in fact been tested by a detailed quantitative analysis
of pion-pion scattering, the process most accurately and extensively 
studied using ChPT.

Consider s-wave $\pi\pi$ scattering at very low energy.  The scattering lengths
in the isospin $I = 0,2$ channels, calculated to leading chiral order,
are\cite{Wei66}
\begin{equation}
a_0 = {7m_\pi\over 32\pi f_\pi^2}~~~~,~~~~a_2 = -{m_\pi \over 16\pi f_\pi^2}~~,
\end{equation}
showing that the $\pi\pi$ interaction properly 
vanishes in the chiral limit, $m_{\pi}\rightarrow 0$. The next-to-leading
order introduces one-loop iterations of the leading ${\cal L}^{(2)}$ part
of the effective Lagrangian as well as pieces generated by ${\cal L}^{(4)}$.
At that level enters the renormalized constant $\bar{l}_3$ which
also determines the correction to the leading-order GOR relation:
\begin{equation}
m_{\pi}^2~ =~\stackrel{o}{m}_{\pi}^2 -
\frac{\bar{l}_3}{32\pi^2 f^2}
 \stackrel{o}{m}_{\pi}^4 + ~{\cal O}(\stackrel{o}{m}_{\pi}^6)~~,
 \label{eq:GOR2}
\end{equation}
where
\begin{equation}
\stackrel{o}{m}_{\pi}^2~=~ - \frac{m_u + m_d}{2\,f^2}\langle \bar{\psi} \psi \rangle
\label{eq:GOR3}
\end{equation}
involves the quark mass in leading order. Here $f$ is the
pion decay constant in the chiral limit. 
An accurate determination of the $I = 0$ s-wave $\pi\pi$ scattering
length therefore provides a constraint for $\bar{l}_3$ which in turn
sets a limit for the next-to-leading order correction to the GOR relation.

Such an investigation has been performed\cite{CGL01} using
low-energy $\pi\pi$ phase shifts extracted from the detailed 
final state analysis of the $K \rightarrow \pi\pi + e\nu$ decay. The result, when 
translated into a statement about the non-leading term 
entering (\ref{eq:GOR2}), implies that the difference between
$m_{\pi}^2$ and the leading GOR expression (\ref{eq:GOR3}) is less
than 5 percent. Hence the ``strong condensate'' scenario of spontaneous chiral
symmetry breaking in QCD appears to be confirmed\footnote{One should note, however,
that this conclusion is drawn at the level of QCD with only $N_f = 2$
flavours. Additional corrections may still arise when strange quarks are
taken into account.}. 

\subsection{The Pion in Lattice QCD}

\begin{figure}
       \centerline{\includegraphics[width=5.5cm] {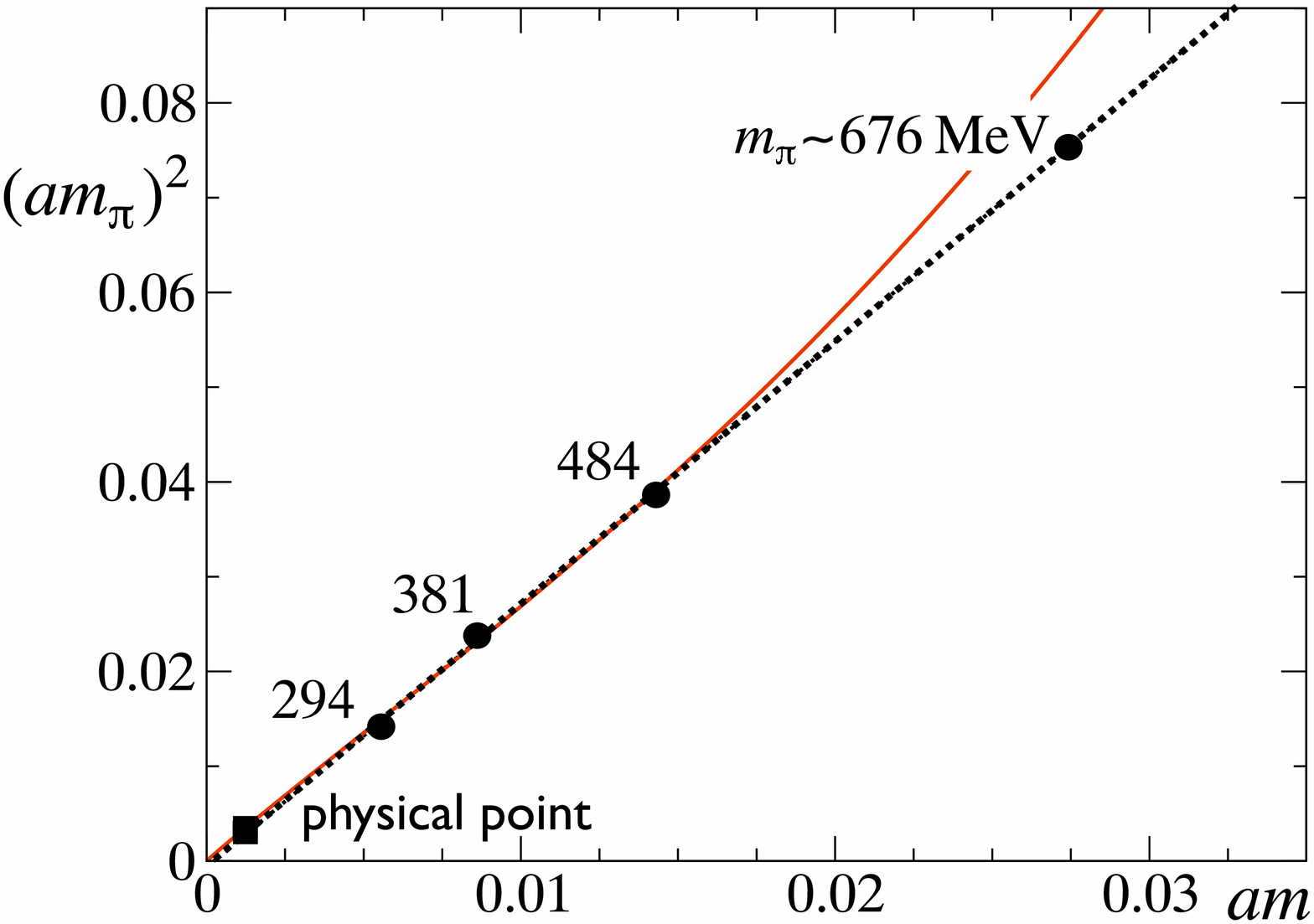}}
   \caption{Lattice QCD simulation results\cite{Lue06}  for the squared pion mass $m_\pi^2$ as function of the quark mass $m$ in units of the lattice spacing $a$. Pion masses converted to physical units are attached to the lattice data points. A linear fit (dashed) is shown in comparison with the next-to-leading order ChPT result (solid curve).}
   \label{fig:4}
 \end{figure}

The leading-order 
relationship $m_\pi^2 \sim m_q$ is also observed\cite{Aoki03}  in lattice QCD
up to surprisingly large quark masses. A detailed recent analysis\cite{Lue06} is shown in Fig.\ref{fig:4}.
Within statistical errors, the data for squared pion mass versus quark mass lie on a straight. The lattice results are remarkably compatible with one-loop chiral perturbation theory up to $m_\pi \lesssim  0.5$ GeV. 

\subsection{Pion-Nucleon Effective Lagrangian}

The prominent role played by the pion as a Goldstone boson of spontaneously
broken chiral symmetry has its impact on the low-energy structure
and dynamics of nucleons as well\cite{TW01}. When probing the nucleon with long-wavelength
electroweak and strong fields, a substantial part of the response comes from the pion
cloud, the ``soft'' surface of the nucleon. 
The calculational framework for this, baryon chiral perturbation theory\cite{EM96,BKM95} 
has been applied quite successfully to a variety of low-energy processes
(such as low-energy pion-nucleon scattering, threshold pion photo- and electroproduction 
and Compton scattering on the nucleon). 

Consider now the sector with baryon number $B = 1$ and the physics of the pion-nucleon system. 
The nucleon is represented by an isospin-$1/2$ doublet, Dirac spinor field $\Psi_N(x) = (p,n)^T$ 
of proton and neutron. The free field
Lagrangian
\begin{equation}
{\cal L}_0^N = \bar{\Psi}_N(i\gamma_{\mu}\partial^{\mu} - M_0)\Psi_N
\end{equation}
includes the nucleon mass in the chiral limit, $M_0$. One should note that the nucleon, 
unlike the pion, has a large mass of the same order as the chiral symmetry breaking scale
$4\pi f_\pi$, which survives in the limit of vanishing bare quark masses, $m_{u,d}\rightarrow 0$. 

The previous pure meson Lagrangian ${\cal L}_{eff}$
is now replaced by ${\cal L}_{eff}(U,\partial^{\mu} U, \Psi_N, ...)$ which also
includes the nucleon field. The additional term involving the nucleon, denoted
by ${\cal L}_{eff}^N$, is expanded again in powers of derivatives (external
momenta) of the Goldstone boson field and of the quark masses: 
\begin{equation}
{\cal L}_{eff}^N = {\cal L}_{\pi N}^{(1)} + {\cal L}_{\pi N}^{(2)} ~...
\end{equation}
In the leading term, ${\cal L}_{\pi N}^{(1)}$ there is
a replacement of $\partial^{\mu}$ by a chiral covariant derivative
which introduces vector current couplings between the pions and the nucleon.
Secondly, there is an axial vector coupling. This structure of the $\pi N$
effective Lagrangian is again dictated by chiral symmetry. We have
\begin{equation}
{\cal L}_{\pi N}^{(1)} =  \bar{\Psi}_N[i\gamma_{\mu}(\partial^{\mu} 
- i{\cal V}^{\mu}) + \gamma_{\mu}\gamma_5\, {\cal A}^{\mu} - M_0]\Psi_N~~,
\label{eq:LeffN}
\end{equation}
with vector and axial vector quantities involving the Goldstone boson (pion)
fields in the form $\xi = \sqrt{U}$:
\begin{eqnarray}
{\cal V}^{\mu} & = & \frac{i}{2}(\xi^{\dagger}\partial^{\mu}\xi +  
\xi\partial^{\mu}\xi^{\dagger}) = -\frac{1}{4f_\pi^2}\varepsilon_{abc}\tau_a\,\pi_b
\,\partial^{\mu}\pi_c + ...~~, \\
{\cal A}^{\mu} & = & \frac{i}{2}(\xi^{\dagger}\partial^{\mu}\xi -  
\xi\partial^{\mu}\xi^{\dagger}) = -\frac{1}{2f_\pi^2}\tau_a\,
\partial^{\mu}\pi_a + ...~~,
\end{eqnarray}
where the last steps result when expanding ${\cal V}^{\mu}$ and ${\cal A}^{\mu}$ to 
leading order in the pion fields. 
So far, the only parameters that enter are the nucleon mass, $M_0$,
and the pion decay constant, $f_\pi$, both taken in the chiral limit.

The nucleon has its own intrinsic structure which leads to a modification of the axial vector
coupling term in (\ref{eq:LeffN}). The analysis of
neutron beta decay reveals that the 
$\gamma_{\mu}\gamma_5$ term is to be multiplied by the axial vector
coupling constant $g_A$, with the empirical value $g_A \simeq1.27$.

At next-to-leading order $({\cal L}_{\pi N}^{(2)})$, the symmetry breaking 
quark mass term enters. It has the effect of shifting the nucleon mass from
its value in the chiral limit to the physical one: 
\begin{equation}
M_N = M_0 + \sigma_N~~.
\end{equation}
The sigma term
\begin{equation}
\sigma_N = m_q\frac{\partial M_N}{\partial m_q} =
\langle N | m_q(\bar{u}u + \bar{d}d) |N\rangle
\end{equation}
measures the contribution of the non-vanishing quark mass, $m_q =
\frac{1}{2}(m_u + m_d)$, to the nucleon mass $M_N$. Its empirical value is in the range
$\sigma_N \simeq (45 - 55)$ MeV and has been deduced\cite{GLS91} by a sophisticated 
extrapolation of low-energy  pion-nucleon data
using dispersion relation techniques.
Up to this point, the $\pi N$ effective Lagrangian, expanded to second order
in the pion field, has the form
\begin{eqnarray}
{\cal L}_{eff}^{N} & = & \bar{\Psi}_N(i\gamma_{\mu}\partial^{\mu} 
- M_N)\Psi_N - \frac{g_A}{2f_{\pi}} \bar{\Psi}_N\gamma_{\mu}\gamma_5\, 
\mbox{\boldmath $\tau$}\,\Psi_N\cdot\partial^{\mu}\mbox{\boldmath $\pi$}  \\
                   &   & -\frac{1}{4f_{\pi}^2}
 \bar{\Psi}_N\gamma_{\mu}\, 
\mbox{\boldmath $\tau$}\,\Psi_N\cdot\mbox{\boldmath $\pi$}\times
\partial^{\mu}\mbox{\boldmath $\pi$}
+{\sigma_N\over  2f_\pi^2}\,\bar{\Psi}_N\Psi_N\,\mbox{\boldmath $\pi$}^2
+ ...~~,\nonumber
\end{eqnarray}
where we have not shown a series of additional terms of order 
$(\partial^{\mu} \mbox{\boldmath$\pi$})^2$ included in the complete ${\cal L}_{\pi N}^{(2)}$.
These terms come with further low-energy constants encoding physics at smaller distances and higher energies. These constants need to be fitted to experimental data, e.g. from pion-nucleon scattering. 

The ``effectiveness" of such an effective field theory relies on the proper identification of the active low-energy degrees of freedom. Pion-nucleon scattering is known to be dominated by the p-wave  $\Delta(1232)$ resonance with spin and isospin 3/2. The excitation energy of this resonance, given by the mass difference $\delta M=M_\Delta - M_N$, is not large, just about twice the pion mass. If the physics of the $\Delta(1232)$ is absorbed in low-energy constants of an effective theory that works with pions and nucleons only (as commonly done in heavy-baryon ChPT), the limits of applicabilty of such a theory is clearly narrowed down to an energy-momentum range small compared to $\delta M$. The $B=1$ chiral effective Lagrangian is therefore often extended\cite{HHK97} by incorporating the $\Delta$ isobar as an explicit degree of freedom. 

\section{Chiral Thermodynamics and Goldstone Bosons in Matter}

Before turning to chiral dynamics in nuclear many-body systems, it is instructive to make a brief digression and touch upon more general issues of chiral symmetry  at finite temperature and non-zero baryon density. 

\subsection{The Chiral Order Parameter}

As outlined in the previous sections, the QCD ground state (the vacuum) is characterized by the presence of the strong chiral condensate $\langle\bar{\psi}\psi\rangle$. The light hadrons are quasiparticle excitations of this condensed ground state, with Yukawa's pion playing a very special role as Nambu-Goldstone boson of spontaneously broken chiral symmetry. A key question\cite{WW03} is then the following: how do the basic quantities and scales associated with this symmetry breaking pattern (the chiral condensate, the pion mass and decay constant) evolve with changing thermodynamical conditions (temperature, baryon density)?

\begin{figure}[htb]
\begin{minipage}[t]{65mm}
\includegraphics[width=5.5cm]{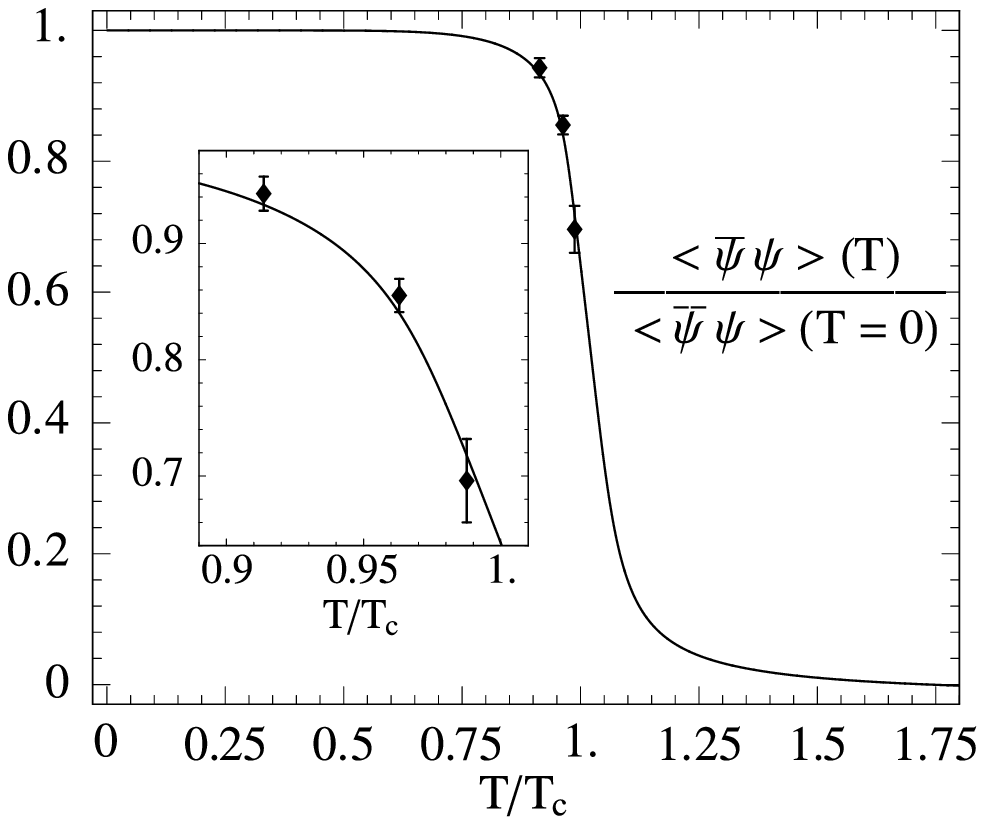}
\caption{Temperature dependence of the chiral condensate at zero chemical potential. The curve results from a calculation\cite{RRW07} based on an extended NJL model with inclusion of Polyakov loop dynamics (the PNJL model). The data points are $N_f = 2$ lattice QCD results taken from ref.\cite{Boyd95} . }
\label{fig:5}
\end{minipage}
\hspace{\fill}
\begin{minipage}[t]{65mm}
\includegraphics[width=7cm]{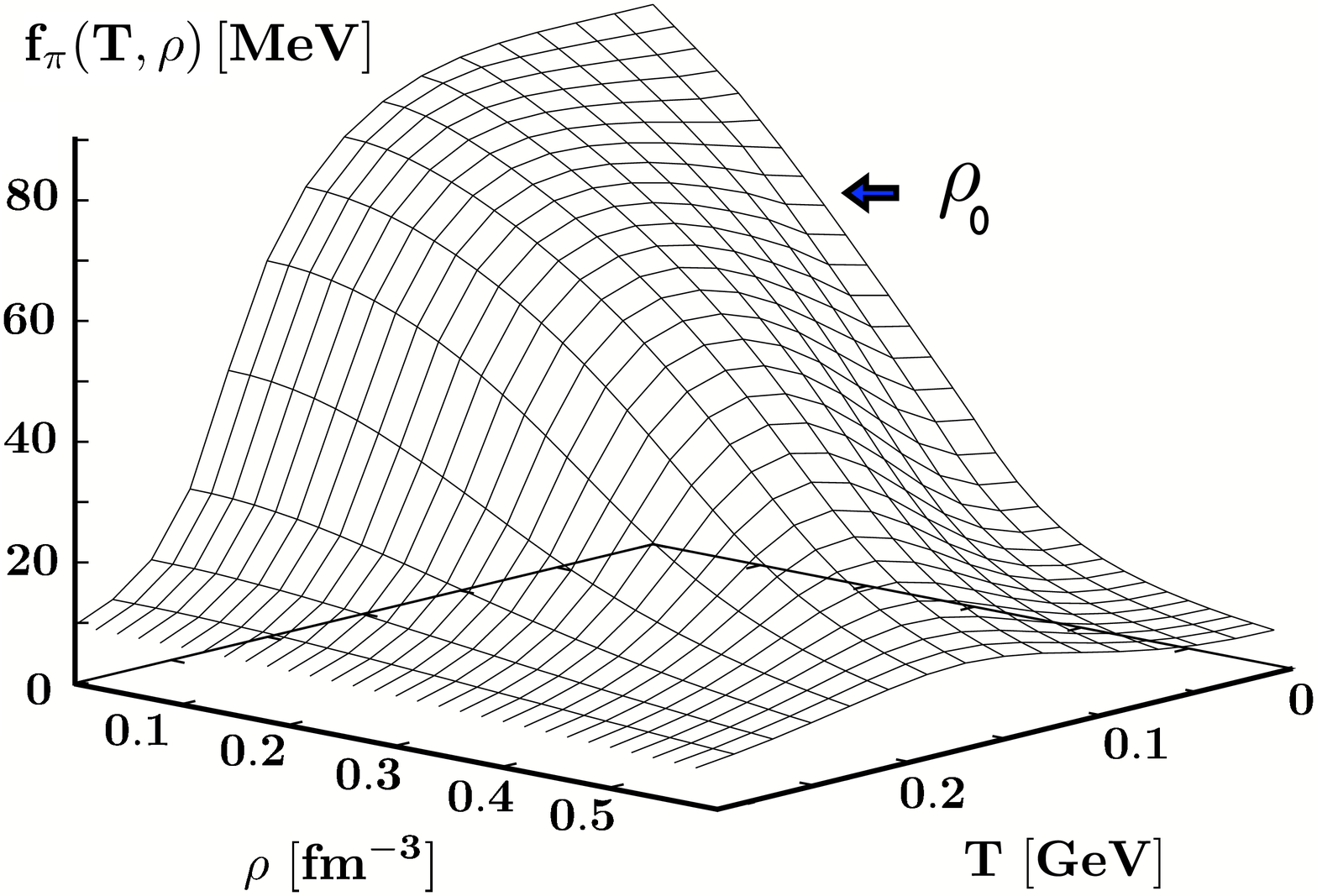}
\caption{Pion decay constant as function of temperature $T$ and baryon density $\rho$. Calculation\cite{RTW06} based on the PNJL model. Normal nuclear matter density $\rho_0 = 0.16$ fm$^{-3}$ is indicated for orientation.}
\label{fig:6}
\end{minipage}
\end{figure}

Assume a homogenous hadronic medium in a volume $V$ at temperature $T$ and consider the pressure
\begin{equation}
P(T,V,\mu) = \frac{T}{V}\ln {\cal Z} = \frac{T}{V}\ln~Tr \exp\left[-\frac{1}{T}\int_V d^3x~({\cal H}-\mu\rho)\right]~~.
\end{equation} 
Here $\mu$ denotes the chemical potential, $\rho$ the baryon density.
The Hamiltonian density ${\cal H}$ of QCD is expressed in terms of the relevant
degrees of freedom in the hadronic phase, derived from the chiral effective Lagrangian
${\cal L}_{eff}$. The $N_f = 2$ Hamiltonian has a mass
term, $\delta{\cal H } = \bar{\psi}m\psi = m_u~\bar{u}u + m_d~\bar{d}d$, so that 
${\cal H} = {\cal H}_0 + \delta{\cal H}$, with ${\cal H}_0$ representing the massless limit. Now take the derivative of the pressure with respect to the quark mass and use the GOR relation (\ref{eq:GOR}) to derive the condensate $\langle\bar{\psi}\psi\rangle_{T,\rho}$ at finite $T$ and density $\rho = \partial P/\partial\mu$, or rather its ratio with the condensate at $T=\mu=0$,
\begin{equation}
\frac{\langle \bar{\psi}\psi\rangle_{T,\rho}}{\langle \bar{\psi}\psi\rangle_0} =
1 + \frac{dP(T,\mu)}{f_{\pi}^2\,dm_{\pi}^2}~~.
\label{con2}
\end{equation}
The $T$ dependence of this condensate, at zero chemical potential, is shown in comparison with two-flavor lattice QCD results in Fig.\ref{fig:5}. Its behaviour reflects a continuous crossover transition at a critical temperature $T_c \sim 0.2$ GeV which turns into a second order phase transition in the chiral limit of massless quarks. Above $T_c$ chiral symmetry is restored and the pion stops being realized as a Nambu-Goldstone mode. The chiral condensate therefore has the features of an order parameter. However, it is not directly observable.  A related measurable quantity is the pion decay constant. Its temperature and density dependence is in fact an indicator of tendencies towards chiral symmetry restoration, in the following sense.   

The GOR relation (\ref{eq:GOR}) continues to hold in matter at finite 
temperature $T < T_c$ and density $\rho$, when reduced to a statement
about the $\it{time}$ component, $A_a^0 = \psi^{\dagger}\gamma_5 (\tau_a/2)
\psi$, of the axial current. We can introduce the in-medium pion 
decay constant, $f_{\pi}^*(T,\rho)$ through the thermal matrix element
$\langle~~|A^0|\pi\rangle_{T,\rho}$ , the in-medium analogue of eq.(\ref{eq:fpi}). 
One finds  
\begin{equation}
f_{\pi}^*(T,\rho)^2\,m_{\pi}^*(T,\rho)^2 = -{m_u + m_d\over 2}
\langle\bar{\psi}\psi\rangle_{T,\rho} + ... ~~
\label{inmedGOR}
\end{equation}
to leading order in the quark mass. The in-medium pion mass
$m_{\pi}^*$ (more precisely: the average of the $\pi^+$ and $\pi^-$
masses) is protected by the pion's Goldstone boson nature and 
not much affected by the thermal environment. The ``melting'' of the condensate by heat or
compression translates primarily to the in-medium change of the pion decay 
constant.

The leading behaviour\cite{GL89,WW03} of  the pion condensate and, consequently,
of the pion decay constant, with increasing temperature and density is:
\begin{equation}
\left({f_\pi^*(T,\rho)\over f_\pi}\right)^2 \sim \frac{\langle \bar{\psi}\psi\rangle_{T,\rho}}{\langle \bar{\psi}\psi\rangle_0} = 
1 - {T^2\over 8\,f_\pi^2} - {\sigma_N\over m_\pi^2\,f_\pi^2}\rho + \, ...~~.
\label{eq:con2}
\end{equation}
A typical result for the in-medium behavior of the pion decay constant is displayed in Fig.\ref{fig:6}.
It should be noted that the dropping of the condensate's magnitude with density is significantly more pronounced than its temperature dependence. The decreasing ``chiral gap'' $4\pi f_{\pi}^*(T,\rho)$ 
with changing thermodynamic conditions should thus imply
observable changes in the low-energy dynamics of pions in dense matter.

\subsection{Low-Energy Pion-Nucleus Interactions}

Goldstone's theorem implies that low-momentum pions interact weakly. 
This is generally true also for low-momentum pions interacting with  
nuclear many-body systems. As a starting point, consider
homogeneous nuclear matter at zero temperature with proton density
$\rho_p$ and neutron density $\rho_n$. A pion wave in matter has its
energy $\omega$ and momentum $\vec{q}$ related by the dispersion
equation  
\begin{equation}
\omega^2 - \vec{q}~^2 - m_{\pi}^2 - 
\Pi(\omega,\vec{q}~;\rho_p,\rho_n) = 0~~~.
\end{equation}
The polarization function, or pion self-energy $\Pi$, summarizes all interactions 
of the pion with the medium. At low densities,
\begin{equation}
\Pi^{(\pm)}(\omega,\vec{q}\,;\rho_p,\rho_n) = -T^{+}(\omega,\vec{q}\,)\,\rho 
\pm T^{-}(\omega,\vec{q}\,)\,\delta\rho~~~,
\label{eq:Pi}
\end{equation}
in terms of the isospin-even $(T^{+})$ and isospin-odd $(T^{-})$ pion-nucleon forward scattering amplitudes, with
$\rho = \rho_p + \rho_n$ and $\delta\rho = \rho_p - \rho_n$.
We have now specified the self-energies $\Pi^{(\pm)}$ for a $\pi^+$
or $\pi^-$, respectively. 

Applications to finite systems, in particular
for low-energy pion-nucleus interactions\cite{EE66,EW88} relevant to pionic atoms, 
commonly make use of an energy-independent effective potential.
Such an equivalent potential is constructed\cite{Mig78} by expanding the
polarization function for $\omega - m_\pi \ll m_\pi$ and $|\vec{q}\,|^2 \ll
m_\pi^2$ around the physical threshold, $\omega = m_\pi$ and $|\vec{q}\,| = 0$.
By comparison with the Klein-Gordon equation for the pion wave function
$\phi(\vec{r}\,)$ in coordinate space,
\begin{eqnarray}\label{kge1}
\left[ \omega^2-m_\pi^2+\vec{\nabla}^2-
2m_\pi U(\vec{r}\,)\right]\phi(\vec r\,)=0\, ~~~,
\end{eqnarray}
the (energy-independent) potential $U(\vec{r}\,)$ is identified
as follows:
\begin{equation}
2m_\pi U(\vec{r}\,) = \left(1-\frac{\partial\Pi}{\partial\omega^2}\right)^{-1}
\left[\Pi(m_\pi,0) -\vec{\nabla}\left(\frac{\partial\Pi}{\partial\omega^2}
+\frac{\partial\Pi}{\partial\vec{q}\,^2}\right)\vec{\nabla}\right]~~~,
\label{eq:pot}
\end{equation} 
with all derivatives taken at the threshold point. The wave 
function renormalization factor $(1-\partial\Pi/\partial\omega^2)^{-1}$ 
encodes the energy dependence of the polarization function
$\Pi(\omega,\vec{q}\,)$ in the equivalent energy-independent potential 
(\ref{eq:pot}). This potential is expressed in terms of local density
distributions $\rho_{p,n}(\vec{r}\,)$ for protons and neutrons, and the
standard prescription $\vec{q}\,^2f(\rho) \rightarrow 
-\vec{\nabla}f(\rho(\vec{r}\,))\vec{\nabla}$ is used for the $\vec{q}\,^2$-
dependent parts. In practical calculations of pionic atoms, the Coulomb
potential $V_c$ is introduced by replacing $\omega \rightarrow \omega -
V_c(\vec{r}\,)$, and corrections of higher order beyond the leading terms
(\ref{eq:Pi}), resulting from double scattering and absorption, are added.  

\subsection{Deeply Bound States of Pionic Atoms}
 
Accurate data on 1s states of a negatively charged pion bound 
to Pb and Sn isotopes\cite{Su04,Ya07} have set new standards and 
constraints for the detailed analysis of s-wave pion interactions with
nuclei. Such deeply bound pionic states owe their existence, with relatively
long lifetimes, to a subtle balance between the attractive Coulomb force
and the repulsive strong $\pi^-$-nucleus interaction in the bulk of
the nucleus. As a consequence, the 1s wave function
of the bound pion is pushed toward the edge of the nuclear surface. Its
overlap with the nuclear density distribution is small, so that the standard
$\pi^- pn \rightarrow nn$ absorption mechanism is strongly suppressed.

The topic of low-energy, s-wave pion-nucleus interactions has a long\cite{EE66,EW88} history.
Inspired by the measurements of deeply bound pionic atoms it has recently been re-investigated\cite{KKW03} from the point of view of the distinct energy dependence of the pion-nuclear polarization
operator in calculations based on systematic in-medium chiral perturbation
theory\cite{KW01,WOM02} . 

Consider a negatively charged pion interacting with nuclear matter and recall the
$\pi^-$ self-energy from Eq.(\ref{eq:Pi}).  
In the long-wavelength limit ($\vec{q}\rightarrow 0$), chiral symmetry
(the Tomozawa-Weinberg low-energy theorem) implies $T^-(\omega) =
\omega/(2 f_{\pi}^2) + {\cal O}(\omega^3)$. Together with the observed
approximate vanishing of the isospin-even threshold amplitude 
$T^+(\omega = m_{\pi})$, it is clear that $1s$ states of pions bound to
heavy, neutron rich nuclei are a sensitive source of information
for in-medium chiral dynamics.  Terms of non-leading order in density (double
scattering (Pauli) corrections of order $\rho^{4/3}$, absorption
effects of order $\rho^2$ etc.) are important and systematically incorporated.  Absorption
effects and corresponding dispersive corrections appear at the
three-loop level and through  short-distance dynamics
parametrized by $\pi N N$ contact terms, not explicitly calculable
within the effective low-energy theory. The imaginary parts
associated with these terms are well constrained  by the
systematics  of observed  widths of pionic atom levels throughout
the periodic table. 

\begin{figure}
\centerline{\includegraphics[width=5.5cm] {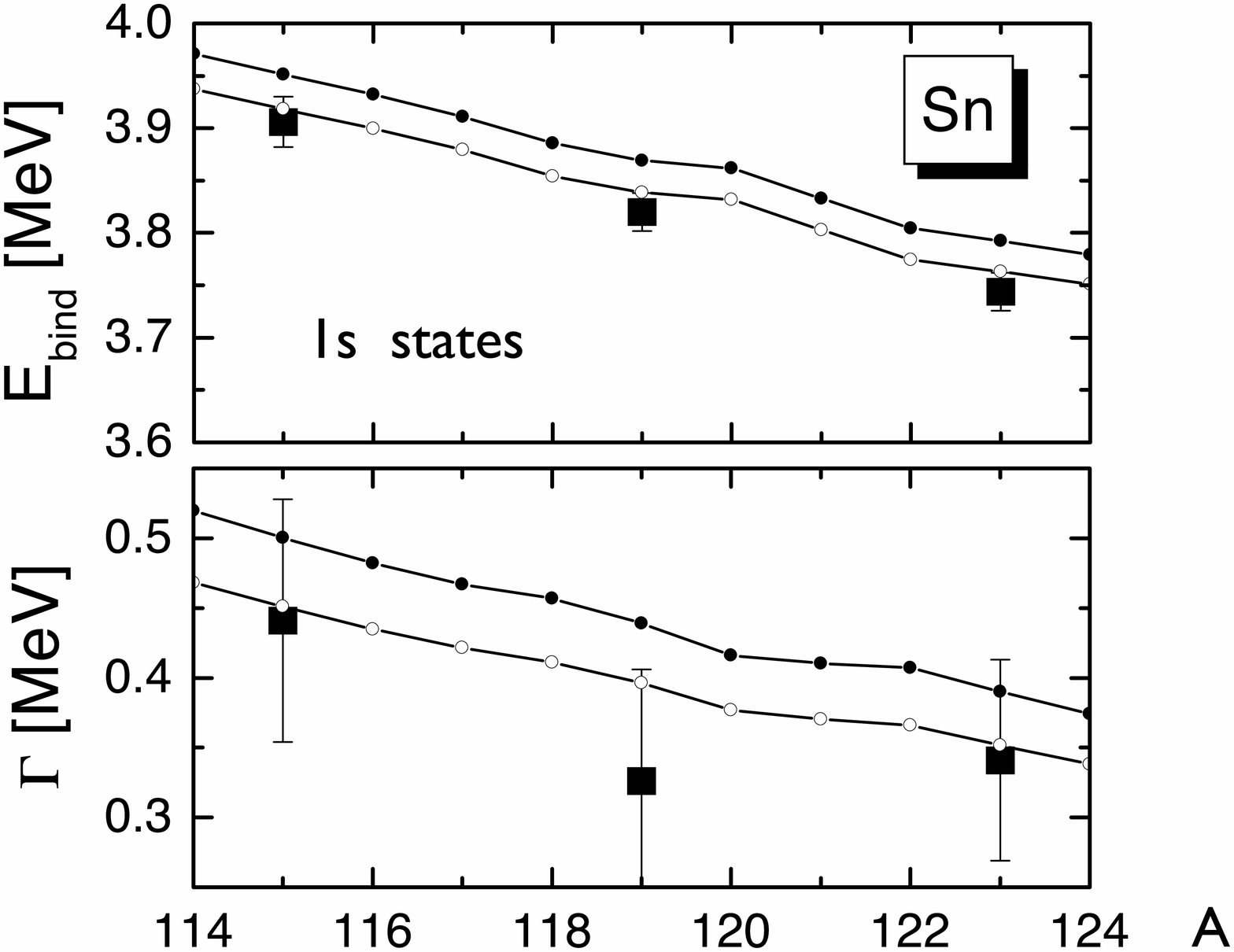}}
\caption{
Binding energies (upper pannel) and widths (lower pannel) of pionic $1s$ states 
in Sn isotopes. The curves show predictions\cite{KKW03} based on the explicitly  
energy dependent pionic s-wave polarization operator calculated in two-loop
in-medium chiral perturbation theory. Upper and lower curves give an impression of 
uncertainties  related to the $\pi N$ sigma term. Data from ref.\cite{Su04}.
}
\label{fig:7}
\end{figure}

With these ingredients the Klein-Gordon equation for the nuclear pion field 
has been solved with the explicitly energy dependent pion
self-energy just described. As an example we show predictions\cite{KKW03}
for binding energies and widths for pionic $1s$ states bound
to a series of Sn isotopes. These calculations
include  a careful assessment of uncertainties in neutron
distributions. Results are shown in Fig.\ref{fig:7} in
comparison with experimental data\cite{Su04}.

The question has been raised\cite{Wei01,KY01} whether one can actually 
"observe" fingerprints of (partial) chiral symmetry
restoration in the high-precision data of deeply bound pionic
atoms. Pionic atom calculations are usually done with \emph{energy independent}
phenomenological optical potentials instead of explicitly energy
dependent pionic polarisation functions. The connection is provided by
Eq.(\ref{eq:pot}). Consider a zero momentum $\pi^-$ in low density 
matter. Its energy dependent leading-order polarisation operator 
is $\Pi(\omega)=-
T^+(\omega)\,(\rho_p+\rho_n)+T^-(\omega)\,(\rho_n - \rho_p)$\,, and the
in-medium dispersion equation at  $\vec {q}=0$ is
$\omega^2-m_\pi^2-\Pi(\omega)=0$\,. The chiral low-energy
expansion of the off-shell amplitude $T^+(\omega)$ at $\vec {q}=0$
implies leading terms of the form  $T^+(\omega)=(\sigma_N-\beta\, \omega^2)/f_\pi^2$,
with $\beta\simeq \sigma_N/m_\pi^2$ required to yield the empirical $T^+(\omega=m_\pi)\simeq 0$.
Using Eq.(\ref{eq:pot}) one finds for the effective (energy-independent) s-wave potential 
\begin{eqnarray}
\label{us2}
U_{\rm S}\simeq \frac{\rho_n-\rho_p}{4\, f_\pi^2}\,
\left(1-\frac{\sigma_N\, \rho}{m_\pi^2\, f_\pi^2}\right)^{-1} =
 \frac{\rho_n-\rho_p}{4\, f_\pi^{* 2}(\rho)}\,,
 \label{eq:Us}
\end{eqnarray}
with the replacement $f_\pi\to f_\pi^*(\rho)$ of the pion decay
constant representing the in-medium wave function renormalization.
The expression (\ref{us2}) is just the one proposed previously in
ref.\cite{Wei01} based on the relation (\ref{eq:con2}) between the
in-medium changes of the chiral condensate 
and the pion decay constant associated with the time
component of the axial current. The explicitly energy dependent
chiral dynamics represented by $\Pi(\omega)$ "knows" about these
renormalization effects. Their  translation into an equivalent,
energy-independent potential implies  $f_\pi\to f_\pi^*(\rho)$ as
given in eq.~(\ref{us2}). This heuristic reasoning 
has recently been underlined by a more profound derivation in ref.{\cite{JHK07} .

The analysis of the deeply bound pionic atom data\cite{Su04} along these lines comes to
the conclusion that, when extrapolated to nuclear matter density $\rho_0 = 0.16$ fm$^{-3}$,
\begin{equation}
f^*_\pi(\rho_0) \simeq 0.8\,f_\pi ~~,
\end{equation}
which is compatible with the theoretical prediction
\begin{equation}
{f^*_\pi(\rho_0)\over f_\pi}  \simeq 1-{\sigma_N\over 2\,m_\pi^2\,f_\pi^2}\,\rho_0~~,
\end{equation}
assuming  $\sigma_N \simeq 50$ MeV.
It is quite remarkable that an optical potential fit to recent precision measurements of $\pi^+$ and $\pi^-$ differential cross sections at the lowest possible energy ($T_\pi = 21.5$ MeV) on a variety of nuclei reaches a similar conclusion\cite{Fried04}, namely $f^*_\pi(\rho_0) \simeq 0.83\,f_\pi$, although within a different procedure. With the interpretation (\ref{eq:Us}), the tendency towards chiral restoration in a nuclear medium as suggested by Eq.(\ref{eq:con2}) appears to be - at least qualitatively - visible in low-energy pion-nucleus interactions. 

\section{Nuclear Chiral Dynamics}

We now approach a basic question at the origin of modern nuclear physics: is there a path from QCD via its low-energy representation, chiral effective field theory, to the observed systematics of the nuclear chart? Or equivalently: how does Yukawa's pion and its realization as a Nambu-Goldstone boson figure in the nuclear many-body problem? 

Pionic degrees of freedom in nuclei have been in the focus right from the beginnings. The field of exchange currents in nuclei, with the pion as the prime agent, was  started already in the early fifties\cite{Mya51} and investigated in great breadth in the seventies. An instructive overview of these developments is given in volume II of ref.\cite{RW79}. The important role played by one-pion exchange and its strong tensor force in the deuteron\cite{ER85} is a long known and well established fact. The description of radiative  $np$ capture ($n+p\rightarrow d+\gamma$) in terms of the magnetic pion exchange current\cite{RB72} was a key to establishing the pion as an observable degree of freedom in the deuteron. 

For heavier nuclei, on the other hand, the role of the pion is not so directly evident.  One-pion exchange does not contribute to the bulk nuclear (Hartree) mean field when averaged over nucleon spins.  Fock exchange terms involving one-pion exchange are relatively small. The role of pions in binding the nucleus is manifest primarily in the intermediate range attractive force generated by two-pion exchange processes. For decades, nuclear mean field models preferred to replace the complexity of such processes by a phenomenological scalar-isoscalar ``sigma" field, although the more detailed treatment of the two-pion exchange nucleon-nucleon interaction had been known before (see e.g. ref.\cite{BJ76} and volume I of ref.\cite{RW79}). Recent developments return to these basics by introducing chiral effective field theory as a systematic framework for the treatment of NN interactions and nuclear systems, following ref.\cite{Wei90}. For an updated review see ref.\cite{Epel06}.  

\subsection{In-medium chiral perturbation theory and nuclear matter}

In nuclear matter the relevant momentum scale is the Fermi momentum $k_F$. Around
the empirical saturation point with $k_F^{(0)} \simeq 0.26$ GeV $ \sim 2m_{\pi}$, the Fermi
momentum and the pion mass are scales of comparable magnitude. This implies that at the densities of 
interest in nuclear physics, $\rho \sim \rho_0 = 2(k_F^{(0)})^3/3\pi^2 \simeq 0.16$ fm$^{-3} \simeq 0.45\, m_\pi^3$, pions must be included as {\it explicit} degrees of freedom: their propagation in matter is "resolved" at the relevant momentum scales around the Fermi momentum.

At the same time, $k_F$ and $m_{\pi}$ are small compared to the characteristic chiral scale, $4 \pi f_{\pi} \sim 1$ GeV. Consequently, methods of chiral perturbation theory are expected to be applicable to nuclear matter at least in a certain window around $k_F^{(0)}$. In that range, the energy density
\begin{equation}
{\cal E}(k_F) = \left[M_N + {E(k_F)\over A}\right]\rho\,.
\end{equation}
should then be given as a convergent power series in the Fermi momentum. This is the working hypothesis. More precisely, the energy per particle has an expansion
\begin{equation}
{E(k_F)\over A} = {3k_F^2\over 10 M_N} + \sum_{n\ge3}{\cal F}_n(k_F/m_\pi)\,k_F^n\,.
\end{equation}
The expansion coefficients ${\cal F}_n$ are in general non-trivial functions of $k_F/ m_{\pi}$, the dimensionless ratio of the two relevant scales. These functions must obviously not be further expanded. Apart from $k_F$ and $m_\pi$, a third relevant ``small" scale is the mass difference $\delta M = M_\Delta - M_N \simeq 0.3$ GeV between the $\Delta(1232)$ and the nucleon. The strong spin-isospin transition from the nucleon to the $\Delta$ isobar is therefore to be included as an additional important ingredient in nuclear many-body calculations, so that the ${\cal F}_n$ become functions of both $k_F / m_\pi$ and $m_\pi / \delta M$.

\begin{figure}[htb]
\begin{minipage}[t]{65mm}
\includegraphics[width=6cm]{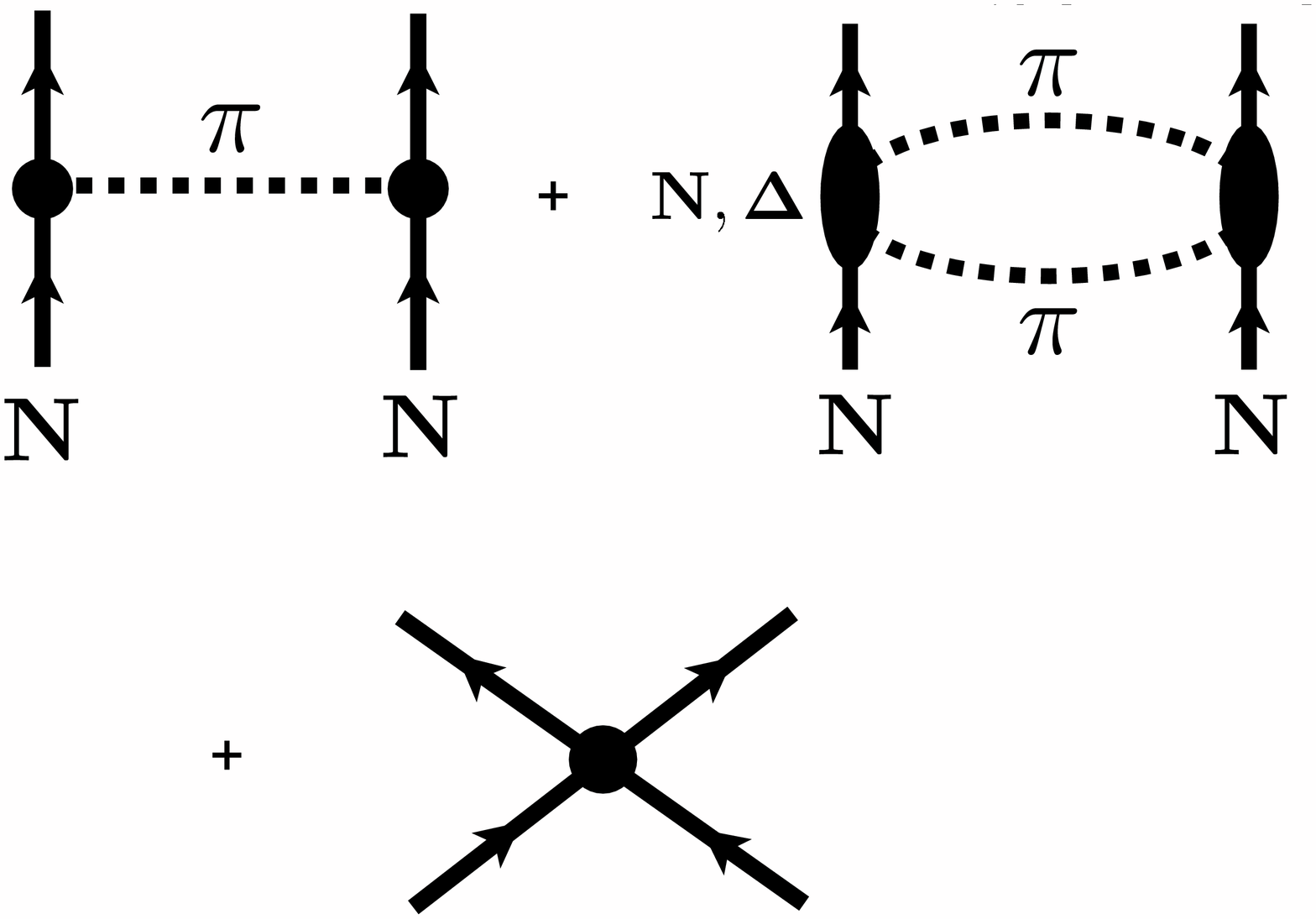}
\caption{NN amplitude in chiral effective field theory:  (upper:) one-pion exchange, two-pion exchange (including $\Delta$ isobar intermediate states) and (lower:) contact terms representing short-distance dynamics.}
\label{fig:8}
\end{minipage}
\hspace{\fill}
\begin{minipage}[t]{65mm}
\includegraphics[width=6.5cm]{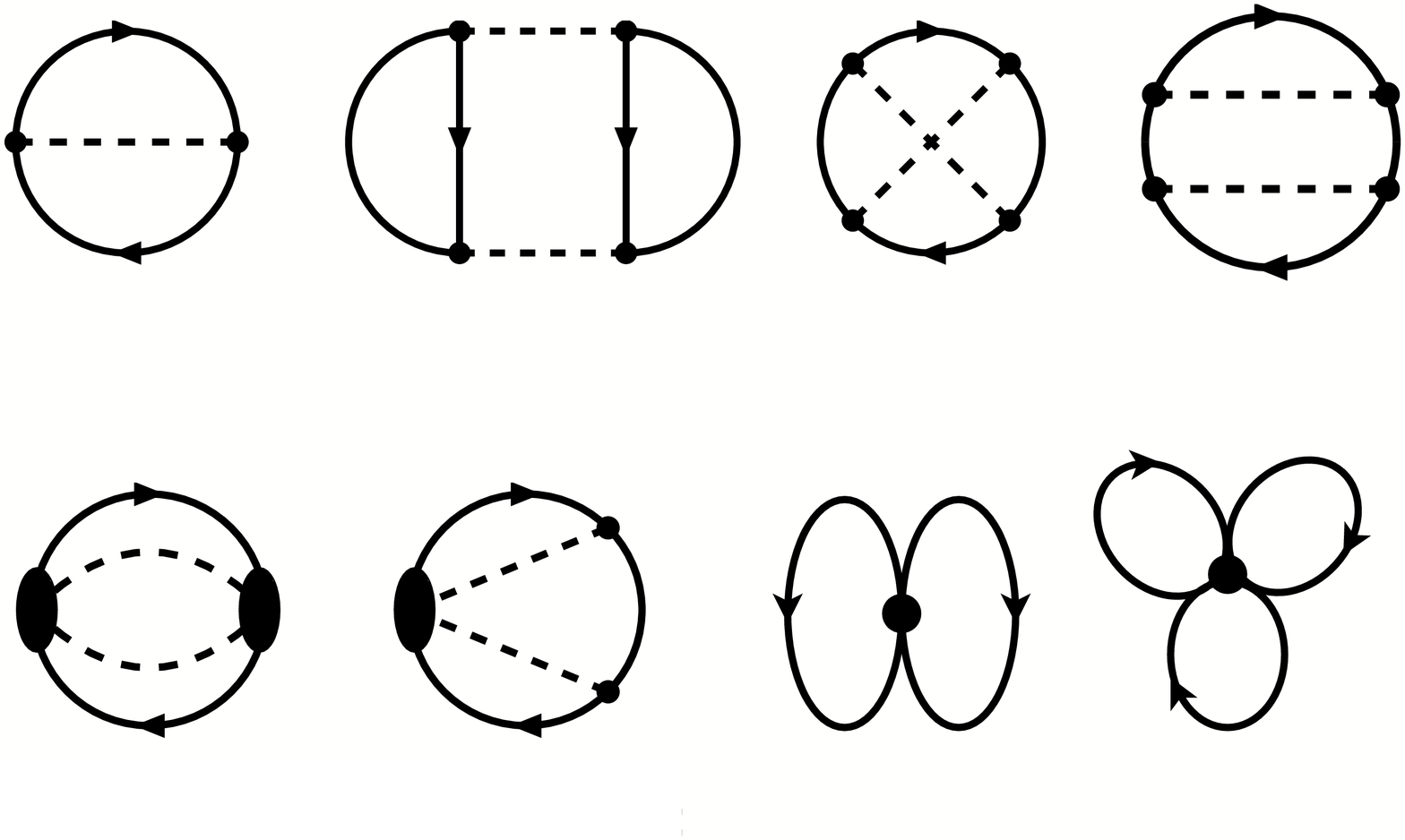}
\caption{Energy density from in-medium chiral perturbation theory at three-loop order. Dashed lines show pions. Each (solid) nucleon line  means insertion of the in-medium propagator (\ref{eq:prop}). Two- and three-body terms involving contact interactions are also shown.}
\label{fig:9}
\end{minipage}
\end{figure}

In-medium chiral perturbation theory is the framework for treating pion exchange processes in the presence of a filled Fermi sea of nucleons. The chiral pion-nucleon effective Lagrangian, with its 
low-energy constants constrained by pion-nucleon scattering observables in vacuum, is used to construct the hierarchy of NN interaction terms as illustrated in Fig.\ref{fig:8}. One- and two-pion exchange processes (as well as those involving low-energy particle-hole excitations) are treated explicitly. They govern the long-range interactions at distance scales $d > 1/k_F$ fm relevant to the nuclear many-body problem, whereas
short-range mechanisms, with t-channel spectral functions involving masses far beyond those of two pions, are not resolved in detail at nuclear Fermi momentum scales and can be subsumed in contact interactions and derivatives thereof. This ``separation of scales" argument makes strategies of chiral effective field theory work even for nuclear problems, with the ``small" scales ($k_F, m_\pi, \delta M$) distinct from the ``large" ones ($4\pi f_\pi, M_N$). In essence, this is the modern realization of Taketani's programme mentioned in the beginning. Closely related renormalization group considerations have motivated the construction of a universal low-momentum NN interaction V($low\, k$)\cite{BKS03} from phase shift equivalent NN potentials such that the ambiguities associated with unresolved short-distance parts disappear. 

The two-pion exchange interaction has as its most prominent pieces the second order tensor force and intermediate $\Delta(1232)$ states which reflect the strong spin-isospin polarizability of the nucleon. The latter produces a Van der Waals - like NN interaction. At long and intermediate distances it behaves as \cite{KBW}
\begin{equation}
V_{2\pi}(r) \sim -{e^{-2m_\pi r}\over r^6}P(m_\pi r)~ ,
\end{equation}
where $P$ is a polynomial in $m_\pi r$. In the chiral limit $(m_\pi \rightarrow 0)$, this $V_{2\pi}$ approaches the characteristic $r^{-6}$ dependence of a non-relativistic Van der Waals potential. 

The two-pion exchange force is the major source of intermediate range attraction that binds nuclei. This is, of course, not a new observation. For example, the important role of the second-order tensor force from iterated pion exchange had been emphasised long ago \cite{KB65}, as well as the close connection of the nuclear force to the strong spin-isospin polarizability of the nucleon \cite{EF81}. The new element that has entered the discussion more recently is the systematics provided by chiral effective field theory in dealing with these phenomena.

With these ingredients, in-medium ChPT calculations of nuclear and neutron matter
have been performed\cite{KFW02,FKW05} up to three-loop order in the energy density, as illustrated in Fig.\ref{fig:9}. Each nucleon line in these diagrams stands for the in-medium propagator
\begin{equation}
(\gamma\cdot p + M_N) \left\{\frac{i}{p^2 - M_N^2 + i \varepsilon} - 2 \pi \delta (p^2
- M_N^2) \theta (p_0) \theta(k_F - | \vec{p}\, |) \right\}~~.
\label{eq:prop}
\end{equation}
The regularization of some divergent loops introduces a scale which is balanced by counter terms (contact interactions) so that the result is independent of this regularization scale\cite{KMW07}. A limited, small number of constants in these contact terms must be adjusted to empirical information such as the
equilibrium density of nuclear matter. Stabilization and saturation of nuclear matter at equilibrium is achieved in a non-trivial and model-independent way: the Pauli principle acting on nucleon intermediate states in two-pion exchange processes produces a repulsive term proportional to $\rho^{4/3}$ in the energy per particle. This partial Pauli blocking counteracts the leading attraction from the term linear in $\rho$. Three-body forces arise necessarily and naturally in this approach. Their contributions is not large at normal nuclear matter density, indicating a convergent hierarchy of terms in powers of the Fermi momentum as long as the baryon density does not exceed about twice the density of equilibrium nuclear matter.  

Binding and saturation of nuclear matter can thus be seen, in this approach, as a combination of phenomena and effects which relate to the names Yukawa, Van der Waals and Pauli. It does then perhaps not come as a surprise that the resulting nuclear matter equation of state, see Fig.\ref{fig:10}, is reminscent of a Van der Waals equation of state. The nuclear liquid turns into a gas at a critical temperature $T_c \simeq 15$ MeV, quite close to the commonly accepted empirical range $T_c \sim 16-18$ MeV.

\begin{figure}
\centerline{\includegraphics[width=8cm] {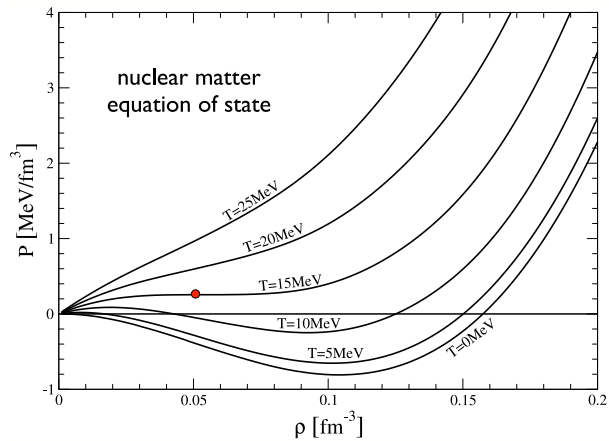}}
\caption{The nuclear matter equation of state: pressure versus baryon density calculated
in three-loop in-medium chiral perturbation theory\cite{FKW05}. Shown are isothermes with temperatures indicated.}
\label{fig:10}
\end{figure}

\subsection{Finite nuclei: density functional strategies}

A description of finite nuclei over a broad range, from $^{16}O$ to the very heavy ones, is  successfully achieved using a (relativistic) universal energy density functional guided by the nuclear matter results. 
The energy as a functional of density is written as:
\begin{equation}
E[\rho] = E_{kin} + \int d^3x\left[{\cal E}^{(0)}(\rho) + {\cal E}_{exc}(\rho)\right] + E_{coul}~~,
\end{equation}
where $E_{kin}$ and $E_{coul}$ are the kinetic and Coulomb energy contributions. 
The basic idea\cite{FKVW04,FKVW06} is to construct ${\cal E}_{exc}$ from the in-medium chiral perturbation theory calculations discussed previously, representing the pionic fluctuations built on the non-perturbative QCD vacuum in the presence of baryons. Binding and saturation, in nuclear matter as well as in finite nuclei, is driven primarily by two-pion exchange mechanisms in combination with the Pauli principle included in ${\cal E}_{exc}$. 

At the same time the QCD vacuum is populated by strong condensates.  The ${\cal E}^{(0)}$ part of the energy density incorporates the leading changes of these condensates at finite baryon chemical potential (or density). As discussed in ref.\cite{FKVW04,FKVW06} and references therein, QCD sum rules at non-zero baryon density suggest that these density dependent changes of condensates generate strong scalar  and vector mean fields with opposite signs: at nuclear bulk densities, several hundred MeV of scalar attraction are compensated by an almost equal amount of vector repulsion, such that the net effect of the condensate mean fields almost vanishes and is hardly visible in infinite, homogeneous nuclear matter. However, In finite nuclei, the coherent effect of the strong scalar and vector mean fields produces the large spin-orbit splitting observed empirically. 

Calculations along these lines, using the chiral pion-nuclear dynamics framework and constraints from the symmetry breaking pattern of low-energy, have been performed in refs\cite{FKVW04,FKVW06} throughout the nuclear chart. The results for nuclear binding energies and radii are comparable in accuracy with those of the best phenomenological relativistic mean field models available. Examples are shown in Figs.\ref{fig:11}, \ref{fig:12} and \ref{fig:13}.   

\begin{figure}
\centerline{\includegraphics[width=6cm] {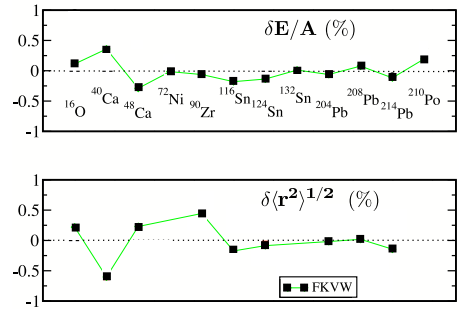}}
\caption{
Deviations (in \%) of calculated binding energies (upper pannel) and r.m.s. charge radii (lower pannel)
from measured values for a series of nuclei from A = 16 to A = 210. For details of the computations see ref.\cite{FKVW06}.
}
\label{fig:11}
\end{figure}

\begin{figure}[htb]
\begin{minipage}[t]{65mm}
\includegraphics[width=6.5cm]{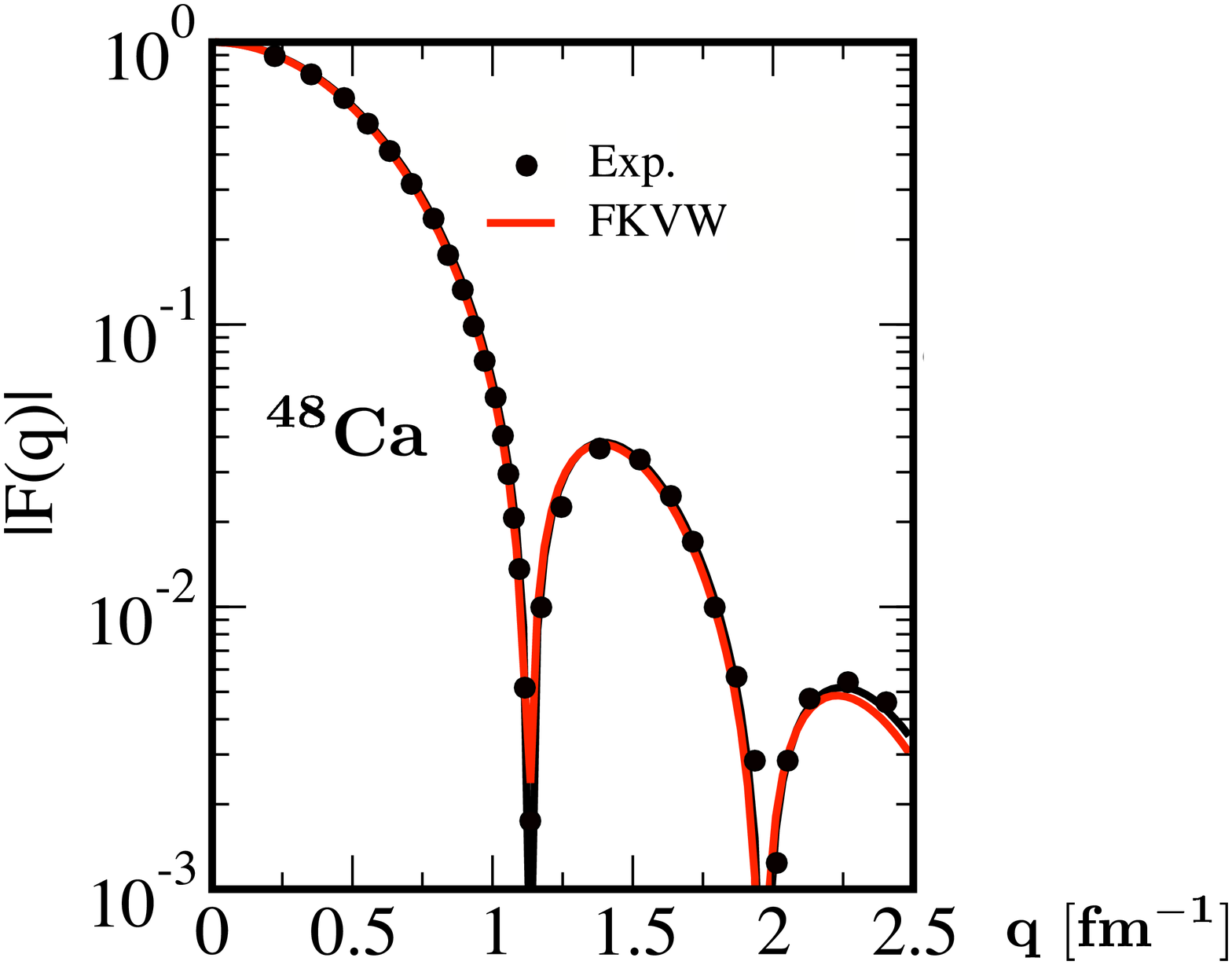}
\caption{Charge form factor of $^{48}$Ca. Calculated curve\cite{FKVW06} in comparison with experimental data.}
\label{fig:12}
\end{minipage}
\hspace{\fill}
\begin{minipage}[t]{65mm}
\includegraphics[width=6.5cm]{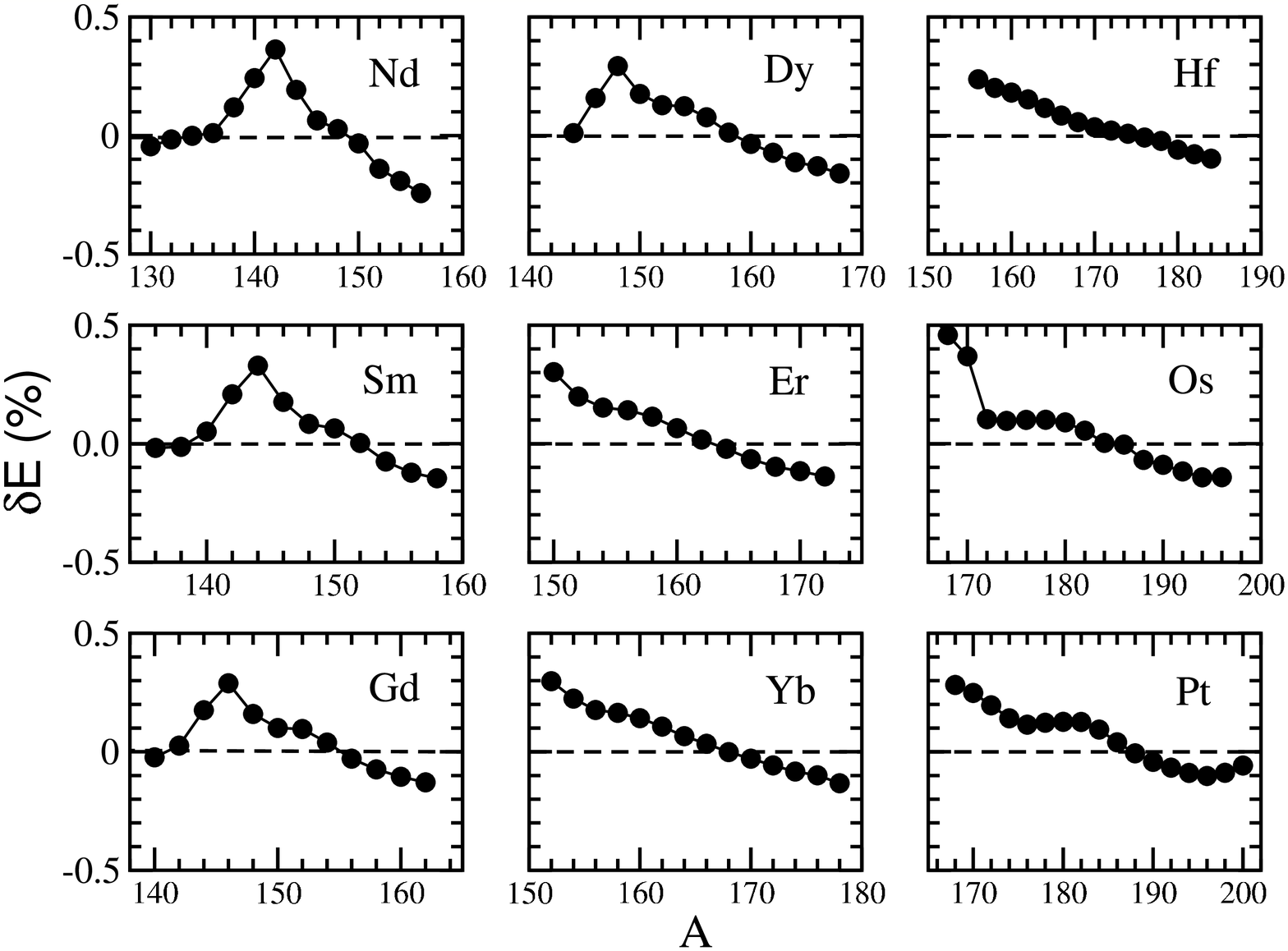}
\caption{Deviations (in \%) of calculated from measured binding energies for a series of isotopic chains from Nd to Pt. For details of the computations see ref.\cite{FKVW06}.}
\label{fig:13}
\end{minipage}
\end{figure}

Of particular interest is the systematics through chains of isotopes of deformed nuclei, increasing the number of neutrons by one unit in each step and changing the deformation pattern along the way. The results are sensitive to the detailed isospin dependence of the nuclear interaction. Chiral pion dynamics and its prediction for the isospin structure of the two-pion exchange NN force in the nuclear medium appears to account successfully for the observed properties of such isotopic chains.

\section{Concluding Remarks}

Yukawa's original U-field which then became the pion is still, more than seventy years after its first release, a generic starting point for our understanding of nuclear systems and interactions. Its property as a Nambu-Goldstone boson of spontaneously broken chiral symmetry is at the origin of a successful effective field theory which represents QCD in its low energy limit. Indications so far are promising that this framework, constrained by the symmetry breaking pattern of low-energy QCD, can serve as foundation for a modern theory of the nucleus. Generating the nucleon-nucleon interaction itself directly from QCD is still a major challenge. Recent lattice QCD results\cite{IAH06}, although still taken at pion masses large compared to the physical one, point to very interesting developments in the near future.

\section*{Acknowledgements} 
It is a great pleasure to thank Professor Taichiro Kugo and his colleagues for their hospitalty in Kyoto
and for arranging a most inspiring Symposium. Avraham Gal's careful reading of the manuscript is gratefully acknowledged.

%


\begin{thebibliography}{99}

\bibitem{Sup55}
Prog. Theor. Phys. Suppl., Volumes {\bf 1} and {\bf 2} (1955). 
  
\bibitem{Yuk35}
 H.~Yukawa, Proceedings of the Physico-Mathematical Society  of Japan   \andvol{17,1935,48}.
 
\bibitem{pion47}
 G. Occhialini, C.F. Powell, C.M.G. Lattes and H. Muirhead, Nature  \andvol{159,1947,186,694}. 

\bibitem{pion48}
 E. Gardner and C.M.G. Lattes, Science  \andvol{107,1948,270}. 

\bibitem{TNS51}
  M. Taketani, S. Nakamura and M. Sasaki, Prog. Theor. Phys.  \andvol{6,1951,581}. 
 
\bibitem{Tak56}
  M.~Taketani, Prog. Theor. Phys. Suppl. \andvol{3,1956,1}. 
  
\bibitem{KMO57}
M. Konuma, H. Miyazawa and S. Otsuki, Phys. Rev. \andvol{107,1957,320}; \\ Prog. Theor. Phys.  \andvol{19,1958,17}.

\bibitem{FM50}
Y. Fujimoto and H. Miyazawa, Prog. Theor. Phys.  \andvol{5,1950,1052}.

\bibitem{G61}
J. Goldstone, Nuovo Cim.  \andvol{19,1961,155}.

\bibitem{NJL61}
Y. Nambu and G. Jona-Lasinio, Phys. Rev.  \andvol{122,1961,345};  \andvol{124,1961,246}.

\bibitem{AD68}
S.L. Adler and R.F. Dashen, {\it Current Algebras}, Benjamin, New York (1968); B.W. Lee, {\it Chiral Dynamics}, Gordon and Breach, New York (1972).

\bibitem{VW91}
U. Vogl and W. Weise W, Prog. Part. Nucl. Phys.   \andvol{27,1991,195}.

\bibitem{HK94}
T. Hatsuda and T. Kunihiro, Phys. Reports  \andvol{247,1994,221}.

\bibitem{KLVW90}
S. Klimt, M. Lutz, U. Vogl and W. Weise, Nucl. Phys. \andvol{A 516,1990,429}.

\bibitem{tH76}
G. 't Hooft, Phys. Rev. \andvol{D 14,1976,3432}.

\bibitem{GOR68}
M. Gell-Mann, R. Oakes and B. Renner, Phys. Rev. \andvol{122,1968,2195}. 

\bibitem{Wei67}
S. Weinberg, Phys. Rev. Lett.   \andvol{18,1967,188}. 

\bibitem{GL84}
J. Gasser and H. Leutwyler, Ann. Phys. \andvol{158,1984,142}. 

\bibitem{Wei79}
S. Weinberg, Physica \andvol{A 96,1979,327}. 

\bibitem{L94}
H. Leutwyler, Ann. Phys. \andvol{235,1994,165}.

\bibitem{Wei66}
S. Weinberg, Phys. Rev. Lett.   \andvol{17,1966,616}. 

\bibitem{CGL01}
G. Colangelo, J. Gasser and H. Leutwyler, Nucl. Phys. \andvol{B 603,2001,125}. 

\bibitem{Aoki03}
S. Aoki et al., Phys. Rev. \andvol{D 68,2003,054502}. 

\bibitem{Lue06}
M. L\"uscher, hep-lat/0509152, PoS \andvol{LAT2005,2006,002}. 

\bibitem{TW01}
A.W. Thomas and W. Weise, {\it The Structure of the Nucleon}, Wiley-VCH, Berlin (2001).

\bibitem{EM96}
G. Ecker and M. Mojzis, Phys. Lett. \andvol{B 365,1996,312}.

\bibitem{BKM95}
V. Bernard, N. Kaiser and U.-G. Meissner, Int. J. Mod. Phys. \andvol{E 4,1995,193}.

\bibitem{HHK97}
T.R. Hemmert, B.R. Holstein and J. Kambor, Phys. Lett. \andvol{B 395,1997,89}.

\bibitem{RTW06}
C. Ratti, M. Thaler and W. Weise, Phys. Rev. \andvol{D 73,2006,014019}.

\bibitem{RRW07}
S. Roessner, C. Ratti and W. Weise, Phys. Rev. \andvol{D 75,2007,034007}.

\bibitem{Boyd95}
G. Boyd et al., Phys. Lett. \andvol{B 349,1995,70}.

\bibitem{GL89}
P. Gerber and H. Leutwyler, Nucl. Phys. \andvol{B 321,1989,387}.

\bibitem{WW03}
W. Weise, {\it Chiral dynamics and the hadronic phase of QCD}, in: Proc. Int. School of Phys. "Enrico Fermi", Course CLIII, A. Molinari et al. (eds.), IOS Press, Amsterdam (2003).

\bibitem{EE66}
M. Ericson and T.E.O. Ericson, Ann. Phys. \andvol{36,1966,383}.

\bibitem{EW88}
T.E.O. Ericson and W. Weise, {\it Pions and Nuclei}, Clarendon Press, Oxford (1988).

\bibitem{Mig78}
A.B. Migdal, Rev. Mod. Phys. \andvol{50,1978,107}.

\bibitem{GLS91}
J. Gasser, H. Leutwyler and M. Sainio, Phys. Lett. \andvol{B 253,1991,252,260}.

\bibitem{KKW03}
E. Kolomeitsev, N. Kaiser and W. Weise, Phys. Rev. Lett. \andvol{90,2003,092501}; \\
Nucl. Phys.  \andvol{A 721,2003,835}

\bibitem{KW01}
N. Kaiser and W. Weise, Phys. Lett. \andvol{B 512,2001,283}.

\bibitem{WOM02}
A. Wirzba, J. Oller and U.-G. Meissner, Ann. Phys. \andvol{297,2002,27}. 

\bibitem{Su04}
K. Suzuki et al., Phys. Rev. Lett. \andvol{92,2004,072302}. 

\bibitem{Ya07}
T. Yamazaki, {\it these Proceedings}, Prog. Theor. Phys. Suppl. (2007).

\bibitem{Wei01}
W. Weise, Nucl. Phys. \andvol{A 690,2001,98}. 

\bibitem{KY01}
P. Kienle and T. Yamazaki, Phys. Lett. \andvol{B 514,2001,1}. 

\bibitem{JHK07}
D. Jido, T. Hatsuda and T. Kunihiro, Proc. YKIS06,  Prog. Theor. Phys. Suppl. (2007).

\bibitem{Fried04}
E. Friedman et al., Phys. Rev. Lett. \andvol{93,2004,122302}; Phys. Rev. \andvol{C 72,2005,034609};
E. Friedman and A. Gal, Phys. Lett. \andvol{B 578,2004,85}.

\bibitem{Mya51}
H. Miyazawa, Prog. Theor. Phys. \andvol{6,1951,801}.

\bibitem{RW79}
M. Rho and D. Wilkinson, {\it Mesons in Nuclei}, North-Holland, \\Amsterdam (1979).

\bibitem{ER85}
T.E.O. Ericson and M. Rosa-Clot, Ann. Rev. Nucl. Sci. \andvol{35,1985,27}.

\bibitem{RB72}
D.O. Riska and G.E. Brown, Phys. Lett. \andvol{B 38,1972,193}.

\bibitem{BJ76}
G.E. Brown and A.D. Jackson, {\it The Nucleon-Nucleon Interaction}, North-Holland, Amsterdam (1976).

\bibitem{Wei90}
S. Weinberg, Phys. Lett. \andvol{B 251,1990,288}; Nucl. Phys. \andvol{B 363,1991,3}. 

\bibitem{Epel06}
E. Epelbaum, Prog. Part. Nucl. Phys. \andvol{57,2006,654}.

\bibitem{BKS03}
S.K. Bogner, T.T.S. Kuo and A. Schwenk, Phys. Reports \andvol{386,2003,1}.

\bibitem{KBW} 
N. Kaiser, S. Gerstend\"orfer and W. Weise, Nucl. Phys. A 637 (1998) 395. 

\bibitem{KB65}
T.T.S. Kuo and G.E. Brown, Phys. Lett. 18 (1965) 54; G.E. Brown, {\it Unified Theory of Nuclear Models and Forces}, 3rd ed., North-Holland, Amsterdam (1971).

\bibitem{EF81} 
J. Delorme, M. Ericson, A. Figureau and C. Thevenet, Ann. Phys. (N.Y.) 102 (1976) 273; M. Ericson and A. Figureau, J. Phys. G7 (1981) 1197.  

\bibitem{KFW02} 
N. Kaiser, S. Fritsch and W. Weise, Nucl. Phys. \andvol{A 700,2002,343}; \\
S. Fritsch, N. Kaiser and W. Weise, Phys. Lett.  \andvol{B 545,2002,73}. 

\bibitem{FKW05} 
S. Fritsch, N. Kaiser and W. Weise, Nucl. Phys. \andvol{A 750,2005,259}.

\bibitem{KMW07} 
N. Kaiser, M. M\"uhlbauer and W. Weise, Eur. Phys. J. \andvol{A 31,2007,53}.

\bibitem{FKVW04} 
P. Finelli, N. Kaiser, D. Vretenar and W. Weise, Eur. Phys. J.  \andvol{A 17,2003, 573}; \\Nucl. Phys.  \andvol{A 735,2004,449}.  

\bibitem{FKVW06} 
P. Finelli, N. Kaiser, D. Vretenar and W. Weise, Nucl. Phys. \andvol{A 770,2006,1}.

\bibitem{IAH06} 
N. Ishi, S. Aoki and T. Hatsuda, nucl-th/0611096.




















\end{thebibliography}
\end{document}